\documentclass[prb,aps,amssymb,twocolumn]{revtex4}

\usepackage[dvips]{graphicx}

\def\be{\begin{equation}}
\def\ee{\end{equation}}
\def\bea{\begin{eqnarray}}
\def\eea{\end{eqnarray}}
\def\nn{\nonumber\\}
\def\({\left(}
\def\){\right)}
\def\HDM{{\cal H}_{\text{\tiny DM}}}
\def\bfk{\mathbf{k}}
\def\bfq{\mathbf{q}}
\def\bfG{\mathbf{G}}

\def\GRPA{{\mathbb G}_{\text{\tiny RPA}}}
\def\PV{\rm P}

\def\zbar{\bar{z}}

\begin{document}

\title{Finite-temperature perturbation theory for
quasi-one-dimensional spin-1/2 Heisenberg antiferromagnets} 

\author{Marc Bocquet}
\affiliation{Department of Physics, Warwick University, 
Coventry CV4 7AL, UK}
\date{October 18, 2001}

\begin{abstract}
We develop a finite-temperature perturbation theory 
for quasi-one-dimensional quantum spin systems,
in the manner suggested by H.J. Schulz in Phys. Rev. Lett.
{\bf 77}, 2790 (1996)
and use this formalism to study their dynamical response.
The corrections to the random-phase approximation formula
for the dynamical magnetic susceptibility
obtained with this method
involve multi-point correlation functions of the
one-dimensional theory on which the random-phase approximation expansion
is built.
This ``anisotropic'' perturbation theory takes the form of a systematic
high-temperature expansion.
This formalism is first applied to the estimation
of the N\'eel temperature of S=1/2 cubic lattice Heisenberg antiferromagnets.
It is then applied to the compound Cs$_2$CuCl$_4$, a frustrated S=1/2
antiferromagnet with a Dzyaloshinskii-Moriya anisotropy.
Using the next leading order to the random-phase approximation, we determine 
the improved values for the critical temperature and incommensurability.
Despite the non-universal character of these quantities, the calculated
values are different by less than a few percent from
the experimental values for both compounds.
\end{abstract}

\maketitle

\section{Introduction}
Quasi-one-dimensional magnets are notoriously difficult to tackle.
The backbones of those compounds, namely the spin chains,
are by now very well understood, in some cases even by analytical methods.
But until now no natural and efficient framework has been
developed to describe their behaviour when they are coupled by a weak
interchain exchange $J_\perp$.

Useful results have nonetheless been obtained by combining one-dimensional exact
results with a random-phase approximation (RPA) approach to cope with interchain
couplings \cite{Schulz1996,Essler1997}.
Recently such a method has even been applied to frustrated
quasi-one-dimensional systems \cite{Bocquet2001}, yielding sensible
predictions.

From the RPA formalism for the dynamical susceptibility, one can deduce
estimates for non-universal quantities, such as the N\'eel temperature
\cite{Schulz1996},
or the possible incommensurate order developing below the transition
in a frustrated antiferromagnet \cite{Bocquet2001}.
This is made possible by recent progress in the (exact)
determination of spin chain two-point correlation functions
in the low-energy regime.
The RPA formalism together with those exact results
are able to cope with exchange anisotropy, and/or
Dzyaloshinskii-Moriya interaction. 

This approach has been successful in the sense that it yields satisfactory
results when compared to experimental measurement (in some cases
even though the interchain ratio $J_\perp/J_\parallel$ is not small
where $J_\parallel$ is the exchange coupling along the easy axis).
This owes to the fact that on one hand the ratio $T_c / J_\parallel$
is small enough so that the collective one-dimensional excitations
have a significant influence on the physics at the transition temperature
(it is always the case for high enough temperature), and on the other hand
in the case of a cubic lattice, $T_c/J_\perp$  is big enough.
However the RPA is an uncontrolled approximation.

V.Y. Irkhin and A.A. Katanin have calculated corrections to RPA
for spin-1/2 quasi-one-dimensional cubic lattices \cite{Irkhin2000}.
Their calculations owe to T. Moriya's empirical improvement
to the RPA formula for the dynamical
susceptibility \cite{Moriya1985} and it differs notably from what follows.
Their work has found applications in the estimation
of the N\'eel temperature of cubic lattice quasi-one-dimensional
antiferromagnets~: KCuF$_3$, Sr$_2$CuO$_3$ and Ca$_2$CuO$_3$.
Their estimation deviates from the RPA result by $25\%$.

In the following, we will develop a systematic expansion and will embed the
RPA formula for the dynamical susceptibility in it as a natural leading
order approximation.
We will be mainly concerned with lattices made of
S=1/2 Heisenberg spin chains. Yet the formalism equally applies to anisotropic
spin chains \cite{Bocquet2001}.
Such an expansion has been developed by E. Arrigoni \cite{Arrigoni2000}
for the physics of Luttinger liquids.
The main differences with our approach are the following.
Because it is at $T=0$, he resums an infinite proper set of cumulants.
On the contrary, we are at finite temperature and we will use the temperature
as an additional energy scale in the disordered phase.
Because of this energy scale, we need not resum all of the higher cumulants
to get a sensible result.
But we do have to resum temperature-dependent diagrams at the level
of any cumulant (four-point correlation function for the examples
given in this work), because we intend to use
our expansion down to the critical temperature, where those diagrams contribute.

In section \ref{GPT} of this paper, we expose the formal steps leading to an extended
perturbative expression of the dynamical susceptibility, in terms
of a self-energy of the two-spin correlation function.
In section \ref{CFC}, we discuss some of the results and peculiarities of this
expansion. In particular we show that the expansion can be organized
in terms of the number of RPA-dressed propagators indirectly related
to the small parameter $J_\perp/T$. This propagator must be regularized
and we hint at how it can be done.
An integral representation of the first correction is given.
In section \ref{ACL}, we calculate the correction due to
the leading diagram in $J_\perp/T$ and discuss its effect
on the physics of cubic lattices and in particular on KCuF$_3$.
In section \ref{AFL}, we investigate the effect of the same correction
in a much more involved case, a quasi-one-dimensional 
(or quasi-two dimensional depending on the point of view since the interchain
coupling is large) frustrated antiferromagnet with a Dzyaloshinskii-Moriya
interaction~: Cs$_2$CuCl$_4$.


\section{General perturbation theory for quasi-one-dimensional magnets}
\label{GPT}


We consider the general quasi-one-dimensional magnetic Hamiltonian
${\cal H}= {\cal H}_{\parallel}+{\cal H}_{\perp}$, where~:
\bea
{\cal H}_\parallel &=& \sum_{i,j} {J_\parallel }_{\mu,\nu}(i,j)
\, S^\mu_{i} S^\nu_{j} \, , \nn
\quad {\cal H}_\perp &=&\sum_{i,j} {J_\perp}_{\mu,\nu}(i,j)
\, S^\mu_{i} S^\nu_{j} \, . 
\eea
The summation over the spin components is implied whereas the latin indices stand for
the sites of the spins.
The quasi-one-dimensional magnetic crystal can be viewed as a set of spin
chains along which the exchange couplings are supposed to be
dominant.
${\cal H}_{\parallel}$ is then defined as the part of ${\cal H}$ which
connects spins on the same spin chain, whereas   
${\cal H}_{\perp}$ connects by definition spins belonging to different spin chains.
We aim at giving a systematic perturbative expansion of the
finite-temperature generating functional~:
\be
Z[\vec{\psi}]={\rm Tr} \left[ T_{\tau} \exp \( -\int_0^\beta \! d\tau \, {\cal H}
- \int_0^\beta \! d\tau \sum_i \vec{\psi}_i.\vec{S}_i \) \right] \, ,
\ee
where $\beta=1/k_B T$. Now we define the isolated spin chain finite-temperature
generating functional~:
\be
Z_{\parallel}[\vec{\psi}] =
{\rm Tr} \left[ T_{\tau} \exp \( -\int_0^\beta \! d\tau \,
{\cal H}_{\parallel} - \int_0^\beta \! d\tau \! \sum_i \vec{\psi}_i.\vec{S}_i
\) \right] \, .
\ee
If we denote $Z_{\parallel}=Z_{\parallel}[\vec{0}]$, then the average of 
the observable $\cal O$ with respect to this functional is
\be
\langle \, {\cal O}[S_i^\mu] \, \rangle_{\parallel} = \frac{1}{Z_{\parallel}}
{\cal O}\left[ \frac{\delta}{\delta\, \psi^\mu_i}\right]
Z_{\parallel}[\vec{\psi}] \, .
\ee
With those notations, we have~:
\be
Z[\vec{\psi}]=Z_{\parallel}\, \langle \exp \( -\int_0^\beta \! d\tau \, {\cal H}_\perp
- \int_0^\beta \! d\tau \! \sum_i \vec{\psi}_i \cdot \vec{S}_i \) \rangle_{\parallel}\, .
\ee
In a very similar fashion as was done in \cite{Boies1995} for coupled Luttinger liquids,
we now introduce a vector field $\vec{\phi}_i(\tau)$ in order to perform
a Hubbard-Stratonovitch transform on ${\cal H}_\perp$~:
\bea
\label{generatrice_1}
Z[\vec{\psi}] &=& Z_{\parallel} \int\! {\cal D} \vec{\phi} \,
\exp \( \frac{1}{4} \int_0^\beta \! d\tau \, \sum_{i,j}
\left[ J_\perp^{-1}\right]_{\mu,\nu}(i,j) \, \phi^\mu_i \phi^\nu_j \) \nn
&& \times \, \langle \exp \( -\int_0^\beta \! d\tau
\sum_i (\vec{\psi}_i+\vec{\phi}_i) \cdot \vec{S}_i \) \rangle_{\parallel}\, .
\eea
The functional integration on $\vec{\phi}$ corresponds to an inverse Laplace
transform.
The second part of the integrand, which is $Z_{\parallel}[\vec{\psi}+\vec{\phi}]$
corresponds to a generating functional of the one-dimensional theory with current source
$\vec{\psi}+\vec{\phi}$. Then $-\ln Z_{\parallel}[\vec{\psi}+\vec{\phi}]$
is the free energy of the sum of the individual
spin chains~: $\ln Z_{\parallel}[\vec{\psi}+\vec{\phi}]=  W[\vec{\phi}+\vec{\psi}]$.
The summation over the spin chains is included in
the functional $W$, which has a Ginzburg-Landau expansion~: 
\be
W[\vec{\phi}] = \frac{1}{2}\int \! d(1) d(2) \, C^{(2)}_{\mu,\nu}(1,2) \,
\phi^{\mu}_{(1)} \phi^{\nu}_{(2)} + W_{\rm I}[\vec{\phi}] \, ,
\ee
where $W_{\rm I}[\vec{\phi}]$ is the interaction functional~:
\bea
W_{\rm I}[\vec{\phi}]&=&\frac{1}{4\,!}\int \! \prod_{i=1}^{4} d(i) \,
C^{(4)}_{\mu,\nu,\lambda,\kappa}(1,2,3,4)
\phi^{\mu}_{(1)} \phi^{\nu}_{(2)} \phi^{\lambda}_{(3)} \phi^{\kappa}_{(4)} \nn
&& + \, O(|\vec{\phi}|^6) \, ,
\eea
where $\int \! d(i)=\int_0^\beta \! d\tau_{i} \int_{-\infty}^{\infty} \! dx \sum_n$
with n the index of the spin chain. $C^{(p)}(1,..,p)$ is the time-ordered imaginary-time
p-point correlation function of an isolated spin chain.

We now work in momentum space and Fourier transform the functional integrals.
We therefore adopt the new convention~:
$\int \! d(i)=\beta \sum_n \int_{-\infty}^{\infty} \! \frac{d k_x}{2\pi}
\int_{0}^{2\pi} \! \frac{d k_y}{2\pi}\int_{0}^{2\pi} \! \frac{d k_z}{2\pi}$
in the case of a three-dimensional magnet.
The summation indexed by $n$  is performed over the Matsubara frequencies
$\omega_n=2\pi n/\beta$. Finally let us define the field theory~:
\be
\label{theory}
\langle {\cal O}\rangle_{\rm I}  = 
{\int\! {\cal D} \vec{\phi} \, \left[ \exp \( F[\phi]\)  \, {\cal O} \right] } /
{\int\! {\cal D} \vec{\phi} \, \exp \( F[\phi]\)} \, ,
\ee
with the weight ($\phi^\alpha_{(1)}$ stands for $\phi^\alpha\,(\omega_1,\vec{k}_1)$)
\bea
\label{theorie_effective}
F[\phi]&=& \frac{1}{2} \int \! d(1) \,
\left[ \left[2\, J_{\perp}\right]^{-1}(1) + C^{(2)}(1)\right]_{\mu,\nu} \,
\phi^\mu_{(1)} \phi^\nu_{(-1)} \nn
&& +  W_{\rm I}[\vec{\phi}] \, .
\eea
Rewriting Eq. (\ref{generatrice_1}) in terms of the theory defined by Eq. (\ref{theory}),
we obtain
\bea
\label{generatrice_2}
\frac{Z[\vec{\psi}]}{Z[\vec{0}]}&=&\exp \( \frac{1}{2} \int \! d(1) \,
\left[2\, J_{\perp}\right]^{-1}_{\mu,\nu}(1)\,
\psi^\mu_{(1)} \psi^\nu_{(-1)}  \) \nn
&& \times \, \langle \, \exp \( -\int \! d(1) \,
\left[2\, J_{\perp}\right]^{-1}_{\mu,\nu} (1) \, 
\psi^\mu_{(1)}\phi^\nu_{(-1)}  \) \, \rangle_{\rm I} \, .
\eea
Interpreting the averaged exponential in (\ref{generatrice_2}) as
a generating functional and introducing the self-energy
$\Sigma_{\mu,\nu}(\omega,\vec{k})$
for the two-point correlation function, we deduce that to second order
in $\vec{\psi}$, one has (assuming for simplicity SU$(2)$ invariance
so that $C^{(2)}_{\mu,\nu}=C^{(2)} \delta_{\mu,\nu}$,
$J^{\mu,\nu}_{\perp}=J_{\perp}\delta_{\mu,\nu}$
and  $\Sigma_{\mu,\nu} =\Sigma \,\delta_{\mu,\nu}$)
\bea
\label{landau-ginzburg}
&&\ln \( Z[\vec{\psi}] / Z[\vec{0}]\) = O(|\vec{\psi}|^4) + \nn
&&\frac{1}{2} \int \! d(1) \, \frac{ C^{(2)}(1) + \Sigma(1)}
{1+2J_{\perp}(1) \( C^{(2)}(1)+\Sigma(1) \)} \,
\vec{\psi}_{(1)}\cdot \vec{\psi}_{(-1)}
\eea
This form of the two-point correlation function has been suggested by
H.J. Schulz in \cite{Schulz1996}.


\section{Calculation of the first corrections to the RPA dynamical susceptibility}
\label{CFC}

\subsection{RPA formula for the dynamical susceptibility}

To the lowest order of approximation, one can set $\Sigma(\omega,\vec{k})=0$
in Eq. (\ref{landau-ginzburg}). We can then continue analytically
(on the frequencies) the one-dimensional two-point correlation function
and therefore recover the dynamical magnetic susceptibility
\be
\chi_{\rm 3d}(\omega,\vec{k})=
\frac{ \chi_{\rm 1d}(\omega,k_x)}
{1-2\,J_{\perp}(\vec{k}) \,
\chi_{\rm 1d}(\omega,k_x)} \, .
\ee
As a consequence, the RPA approximation for quantum spin systems
appears as the leading order of a more general expansion scheme.

\subsection{Higher-order corrections to the RPA formula}

This clearly shows that a systematic expansion can be used.
The free (Euclidian) propagator of the effective theory is an RPA-dressed propagator
(simply called $G$ thereafter).
Its inverse can be read off from Eq. (\ref{theorie_effective})~:
\be
\label{propagateur_RPA}
\left[G \right]^{-1}=C^{(2)}+ \left[2\, J_{\perp} \right]^{-1} \, .
\ee
In real space and imaginary-time, it is given by~:
\bea
\label{propagateur_RPA_2}
G(\tau,\vec{r})&=& \frac{1}{\beta} \sum_n \int_{-\infty}^{\infty} \!
\frac{dk_x}{2\pi} \int_0^{2\pi} \!
\frac{dk_y}{2\pi}\frac{dk_z}{2\pi} \, e^{i\vec{k}\cdot \vec{r}+i\omega_n \tau} \nn
&&\times \frac{2J_{\perp}(\vec{k})}{1+2J_{\perp}\, (\vec{k})
\, C^{(2)}(i\omega_n,\vec{k})} \, .
\eea
Depending on the value of the temperature, the integral might be improper
and it is then meant that the principal value of the integral has to be
taken. We postpone the discussion on this issue to subsection \ref{prescription}.

The vertices of the perturbation theory are given by the
multiple 2n-point correlation functions of the spin chains.
In the case of the spin S=1/2 those are known exactly in the asymptotic limit.

The vertices of the effective field theory (\ref{theorie_effective})
involve separated space-time points $(\tau,x)$
and therefore always depend on $(\omega,k_x)$ when written in momentum space.
On the other hand, they are point-like vertices as far as the transverse space
coordinates are concerned or, stated differently,
do not depend on $\vec{k}_\perp$ in momentum space.
As a consequence, all diagrams in the expansion of the self-energy
$\Sigma(\omega,k_x,\vec{k}_\perp)$ are expected to depend on $k_x$.
Yet, only those with internal RPA-dressed propagator lines which are true functions
of the input transverse momentum $\vec{k}_\perp$ are to depend on it.

\begin{figure}[h]
\includegraphics[width=7cm]{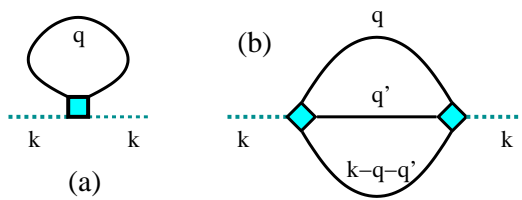} 
\caption{\label{fig:diagram} Diagram (a) does not depend on $\vec{k}_\perp$ 
whereas diagram (b) containing RPA internal lines depending on
the input $\vec{k}_\perp$ does.}  
\end{figure}

The first $\vec{k}_\perp$-dependent diagram possesses three RPA-dressed
propagators and two four-point vertices as depicted on Fig. (\ref{fig:diagram}).
What happens for the transverse momenta is reminiscent of
what occurs to many-body field theories
of electron gas, where the dependence on the space momenta appears
only to the order of this diagram, whereas the first diagrams (Hartree-Fock)
depend only on frequencies.
For diagrams which do not depend on the transverse momenta, and which
are therefore merely one-dimensional, one can resort to the simplified
one-dimensional RPA-dressed propagator~:
\be
G(\tau,x)= G(\tau, x,\vec{r}_{\perp}=\vec{0})\, .
\ee
It is the propagator which has been used in \cite{Irkhin2000}.

What is the small parameter of this expansion ?
In the simplest case of the cubic lattice, it is likely to be
$J_\perp/T$. More precisely it is $A J_\perp/T$ where $A$ is a prefactor,
possibly weakly dependent on the temperature.
The prefactor $A$ will be given later (Eq. (\ref{A})) in the case of S=1/2 Heisenberg
spin chains.
Indeed, each RPA-line contributes by an obvious factor of $(A J_\perp/T)^2$
in any diagram.
Yet in each RPA-line expression remains a non-polynomial dependence
on $A J_\perp/T$ corresponding to the usual RPA re-summation of transverse
paths.
Undoing this RPA summation, the propagator can be expanded in contributions 
with an exact dependence in $(A J_\perp/T)^{2}$, $(A J_\perp/T)^{3}$, etc.
For a diagram with $p$ RPA-lines in it, it is rather
$(A J_\perp/T)^{2p}$, $(A J_\perp/T)^{2p+1}$, etc.

So whatever the subtle dependence of the RPA-dressed propagator on the
temperature, this expansion can genuinely be seen as a high-temperature expansion
in the parameter $(A J_\perp/T)^2$.
More formally, it is also an expansion in the number
of RPA-dressed propagator lines although their dependence in the small parameter
is more intricate.

As a consequence, the conditions of applicability of this perturbation theory
are that $(A J_\perp/T)^2 \ll 1$ but also $T/J_\parallel \ll 1$ in order
for the field-theoretic tools to be valid (in particular in the calculations
of the spin correlation functions at finite temperature).
For the compounds studied here those conditions (which have to be modified
in the case of a frustrated magnet) turn out to be satisfied.

The expansion also depends on the dimensionality of the lattice. This dependence
is obvious at the order of RPA, where the small parameter is there proportional
to the transverse coordination number (see \cite{Irkhin2000}).
The dependence is far less clear at higher order, where the dimensionality
is encrypted in multidimensional integrals.
It is nevertheless possible to take the $d \rightarrow \infty$ limit in these
integrals in order to study this dependence.
But this is beyond the scope of this work.

\subsection{Details for the first correction}

\begin{figure}[h]
\includegraphics[width=6.5cm]{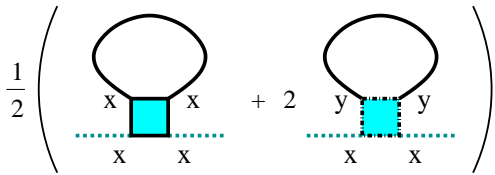} 
\caption{\label{fig:Hartree_Fock} Hartee-Fock diagrams with symmetry factors which
are the first non-trivial terms of the self-energy $\Sigma(\omega,\vec{k})$}  
\end{figure}

Let us take into account the very first correction to the dynamical susceptibility.
So we consider the first non-trivial term in the perturbative
expansion of the self-energy.
It involves the four-point correlation functions
of the spin S=1/2 Heisenberg spin chain.
We decide to truncate the Landau-Ginzburg expansion of $W_{\rm I}[\vec{\phi}]$
to the quartic term in $\vec{\psi}$
(six- and higher-point correlation function do not contribute at this order anyway).
The field theory expansion formally resembles
a 3-component $\vec{\phi}^4$ theory.
In particular the very first correction to the self-energy is given by Hartree-Fock
diagrams (Fig. (\ref{fig:Hartree_Fock})).
The ``free'' propagator of this $\vec{\phi}^4$ field theory
Eq. (\ref{propagateur_RPA_2}) is built on the usual imaginary-time
two-spin correlation function of the Heisenberg chain but dressed by the RPA corrections.
It is therefore a significantly enhanced propagator
and the first correction to RPA in this scheme is expected not to be negligible.

The only vertices of this truncated field theory are given by the four-point
spin correlation functions.
The staggered four-point correlation functions at the isotropic point
can be computed thanks to bosonisation.
We will denote by $A(\Lambda/T)$ the product of the Lukyanov-Zamolodchikov prefactor
with the logarithmic correction induced by the marginally irrelevant
current-current correction \cite{Lukyanov1998,Barzykin2000}:
\be
\label{A}
A\(\frac{\Lambda}{T}\)= \frac{2}{(2\pi)^{3/2}} \, \sqrt{\ln \( \frac{\Lambda}{T}\)
+\frac{1}{2}\ln \ln \( \frac{\Lambda}{T}\)} \, .
\ee
For clarity, we decompose the imaginary-time four-point correlation function
into $A(\Lambda/T)$ and the purely conformal part
of the correlation functions $C^{(4)}_{\rm xxxx}$ and $C^{(4)}_{\rm xxyy}$~:
\bea
T_\tau \langle S^x_{(0)}S^x_{(z_1)}S^x_{(z_2)}S^x_{(z_3)} \rangle & = &
A^2\(\frac{\Lambda}{T}\)C^{(4)}_{\rm xxxx}(z_1,z_2,z_3)\, , \nn
T_\tau \langle S^x_{(0)}S^x_{(z_1)}S^y_{(z_2)}S^y_{(z_3)} \rangle  &=& 
A^2 \(\frac{\Lambda}{T} \) C^{(4)}_{\rm xxyy}(z_1,z_2,z_3)  \, ,
\eea
where $C^{(4)}_{\rm xxxx}$ as well as $C^{(4)}_{\rm xxyy}$ are given by~:
\bea
&& C^{(4)}_{\rm xxxx}(z_1,z_2,z_3)= (-1)^{x_1+x_2+x_3}\,
\times  \nn
&& \left[ \frac{|\Theta(z_1)\Theta(z_2-z_3)|}
{|\Theta(z_2)\Theta(z_3)\Theta(z_1-z_2)\Theta(z_1-z_3)|}
-\frac{2}{|\Theta(z_1)\Theta(z_2-z_3)|} \right. \nn
&& + \frac{|\Theta(z_2)\Theta(z_1-z_3)|}
{|\Theta(z_1)\Theta(z_3)\Theta(z_1-z_2)\Theta(z_2-z_3)|}
-\frac{2}{|\Theta(z_2)\Theta(z_1-z_3)|} \nn
&& \left. + \frac{|\Theta(z_3)\Theta(z_1-z_2)|}
{|\Theta(z_1)\Theta(z_2)\Theta(z_1-z_3)\Theta(z_2-z_3)|} \right. \nn
&& \left. -\frac{2}{|\Theta(z_3)\Theta(z_1-z_2)|} \right]\, , 
\eea
and
\bea
&& C^{(4)}_{\rm xxyy}(z_1,z_2,z_3)= (-1)^{x_1+x_2+x_3}
\frac{1}{|\Theta(z_1)\Theta(z_2-z_3)|} \nn
&& \times \, {\rm Re} \(
\sqrt{\frac{\Theta(z_2)\Theta(\zbar_3)\Theta(\zbar_1-\zbar_2)\Theta(z_1-z_3)}
{\Theta(\zbar_2)\Theta(z_3)\Theta(z_1-z_2)\Theta(\zbar_1-\zbar_3)}}-1 \) \, ,
\eea
where we have denoted
\be
\Theta(z=x+i\tau)=\frac{u}{\pi T}\,\sinh \( \frac{\pi T}{u}(x+iu \tau)\) \, .
\ee
Only the staggered part of the correlation functions which dominate has been
taken into account.

The non-universal constant $\Lambda$ is taken to be $\Lambda=24.27 J_{\parallel}$
as calculated in \cite{Barzykin2001}. Although the value of $\lambda$ and of $\Lambda$
are somehow different from those extracted from numerics \cite{Takigawa1996}
and used in \cite{Irkhin2000}, there is no contradiction  with the
numerical estimates of the correlation functions themselves \cite{Barzykin2001}.
In particular, the N\'eel temperature estimated through RPA for KCuF$_3$
is very close. Discrepancies might nevertheless appear for a different range of
temperature and when the self-energy corrections are taken into account.
Finally $u=\frac{\pi}{2} J_{\parallel}$ is the spin-1/2 Heisenberg chain velocity
given by Bethe Ansatz.

The first contribution to the self-energy is then~:
\begin{widetext}
\be
\Sigma^{(1)}(\omega,\vec{k}) = \frac{1}{2} A^2\(\frac{\Lambda}{T}\)
\int_0^{\beta} \! d\tau_1 d\tau_2 d\tau_3 \int_{-\infty}^{\infty}
\! dx_1 dx_2 dx_3  e^{-ik_x x_1-i\omega \tau_1} G(z_3-z_2)
\left[
\frac{1}{8} \, C^{(4)}_{\rm xxxx}+ \frac{1}{2} \, C_{\rm xxyy}^{(4)}\right]
(z_1,z_2,z_3) \, ,
\ee
\end{widetext}
where the integrals are performed in real space like was done in \cite{Irkhin2000}.
We are mainly interested in the knowledge of $\Sigma(\omega,\vec{k})$
around $k_x=\pi$ because the isotropic
correlation functions are most significant at this point.
More precisely, most of the spectral weight remains at this point.
It turns out that, for our purpose, the numerics are much more efficient
when done in momentum space~:
\bea
\label{sigma_1}
&&\Sigma^{(1)}\(\bfk ,\vec{k}_\perp\)=\frac{1}{2\beta}A^2\(\frac{\Lambda}{T}\)
 \sum_n \int_{-\infty}^{\infty} \! \frac{dq_x}{2\pi} \,
\left[ \frac{1}{8} \, C^{(4)}_{\rm xxxx} \right. \nn
&&\left. + \frac{1}{2} \, C^{(4)}_{\rm xxyy}\right] (\bfk,-\bfk,\bfq_n,-\bfq_n) \,
\int^{2\pi}_{0} \! \frac{d\vec{q}_\perp}{(2\pi)^2}
\, G(\bfq_n,\vec{q}_\perp) \, ,
\eea
where $\bfk=(\omega,k_x)$ and $\bfq_n=(\omega_n,q_x)$.

Now only $C^{(2)}(\omega_n,k_x)$, which enters in $G(\bfq_n,\vec{q}_\perp)$
through Eq. (\ref{propagateur_RPA_2}), remains to be known.
It is the spin-spin time-ordered imaginary-time (isotropic) correlation function.
The large distance behaviour of the finite-temperature
correlation function can be determined by combining results
obtained from the Bethe Ansatz solution \cite{Bethe1931,Korepin1993} of 
the Heisenberg S=1/2 spin chain, with field
theoretic techniques \cite{Luther1975,Schulz1983,Lukyanov1998,Affleck1998,Barzykin2000}
\be
\label{chixxz}
C^{(2)}(\tau,x)=
(-1)^x \, \frac{A(\Lambda/T)}{2}\, 
\frac{{\pi T}/{u}}
{ |\sinh\(\frac{\pi T}{u} (x+iu\tau)\)|} \, .
\ee
This result can be extended to the anisotropic spin chain as well,
although $\Lambda$ is known exactly only at the isotropic point.
The frequency and momentum dependence  is
obtained by Fourier transformation and analytic continuation of the
time-ordered imaginary-time staggered correlation function (\ref{chixxz}) (see
\cite{Schulz1983,Barzykin2001,Schulz1998,Tsvelik1995})~: 
\bea
\chi_{\rm 1d}(\omega,\pi+k_x)&=& -\frac{A(\Lambda/T)}{4\,T}
\frac{ \Gamma\(\frac{1}{4}-i\frac{\omega-uk_x}{4\pi
T}\)}{\Gamma\(\frac{3}{4}-i\frac{\omega-uk_x}{4\pi
T}\)}\nn
&&\times \frac{\Gamma\(\frac{1}{4}-i\frac{\omega+uk_x}{4\pi T}\)}
{\Gamma\(\frac{3}{4}-i\frac{\omega+uk_x}{4\pi T}\)} \, . 
\eea
But we will mostly use this result in its Euclidian form (before
analytic continuation)~:
\be
C^{(2)}(\omega_n,k_x)=-\chi_{\rm 1d}(i\omega_n,k_x) \, .
\ee
Finally let us mention the fact that as claimed before,
$\Sigma^{(1)}(\bfk, \vec{k}_\perp)$ does not actually depend
on $\vec{k}_\perp$. We also note from the above
result that the obvious prefactor
of $J_\perp \Sigma^{(1)}$ is expected to be $(AJ_\perp /T)^3$.

\subsection{Prescription for the RPA-dressed propagator}
\label{prescription}

The RPA-dressed propagator in momentum space (\ref{propagateur_RPA})
may, for some values of the variables ($\omega_n,\vec{k}$) and parameter
($T$), exhibits a singularity. In the explicit formula for the diagrams,
this propagator is always integrated over and the principal
values of the resulting integrals are finite.
Still, the presence of this singularity has to be understood and a correct
prescription for it (here taking the principal value of improper integrals)
to be justified. The difficulty arising from its presence can be overcome
as follows.

Let us assume that this field theory has a critical temperature $T_c$,
which is the exact theoretical estimate of the N\'eel temperature.
The perturbation theory is expected to be valid for high enough temperatures
and only for temperatures above $T_c$ in the disordered phase.
The RPA approach provides an estimated critical temperature $T'_c$
presumably larger (as will be verified later) than $T_c$.
It corresponds to a pole in the dynamical susceptibility.
For $T \geq T'_c$ no singularity is expected to appear in the RPA propagator
and no problem occurs in the perturbation expansion.
Whereas when $T_c \leq T \leq T'_c$, the denominator of the propagator is negative
which is tantamount to realize that subdiagrams given by an RPA line
just add up to infinity.
Nonetheless this is an unphysical singularity which can be cured by a proper
prescription as we are going to hint at.

\begin{figure}[h]
\includegraphics[width=8cm]{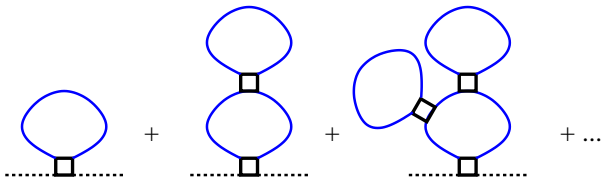} 
\caption{\label{fig:bubbles}Bubbles diagrams that add up to form ${\bf \Sigma}^{(1)}$}  
\end{figure}

Let us show how it is done in the case of the calculation of $T''_c$, the
critical temperature at the next leading order like was done above
when calculating $\Sigma^{(1)}$.
We may expect $T''_c $ to satisfy $T''_c \leq T'_c$ and hence yield a problem.
This statement is based on the fact that including the effects of
the four-spin correlation functions (in addition to
the two-spin correlation functions) amounts to take into account quantum
fluctuations more precisely, as compared to merely restricting to
the Gaussian fluctuations of the RPA approximation. 
As we have seen $\Sigma^{(1)}$ depends on a singular RPA-dressed propagator.
Now, we will add to $\Sigma^{(1)}$ (which sum will hence be denoted ${\bf \Sigma}^{(1)}$)
subdiagrams which would normally be appearing at higher
order in the expansion.
The single RPA line drawn in the diagram for $\Sigma^{(1)}$ appears now as a skeleton line
in the diagram for ${\bf \Sigma}^{(1)}$.
In this case, it stands for an RPA line plus the self-energy correction ${\bf \Sigma}^{(1)}$
itself.
${\bf \Sigma}^{(1)}$ is therefore defined as the sum of all irreducible bubble diagrams
as drawn in Fig. (\ref{fig:bubbles}).
It satisfies a self-consistency Dyson equation given
by the expression (\ref{sigma_1}) but where $G(\bfq_n,\vec{q}_\perp)$ is now replaced by
${\bfG}(\bfq_n,\vec{q}_\perp)$ defined by
an enhanced version of Eq. (\ref{propagateur_RPA})
\be
\left[ {\bfG} \right]^{-1}=C^{(2)}+ \left[2\, J_{\perp} \right]^{-1}
+{\bf \Sigma}^{(1)} \, .
\ee
By construction, the singularity is now avoided, the pole of ${\bfG}$
being displaced to a boundary of the integration domain.
At this boundary the singularity is genuinely integrable
provided the lattice is at least three-dimensional (which is fully
consistent with the Mermin-Wagner theorem).
Hence subdiagrams have ultimately canceled the singularity encountered above.
Yet our method is a systematic one and does not rely on a self-consistent
approach. But we can now extract from this construction the part of
${\bf \Sigma}^{(1)}$ which corresponds to the regularized $\Sigma^{(1)}$
we have to calculate in our scheme.
In order to do so, we decompose the integral 
$\int^{2\pi}_{0} \! \frac{d\vec{q}_\perp}{(2\pi)^2}
\, {\bfG} (\bfq_n,\vec{q}_\perp)$ appearing in the expression of ${\bf \Sigma}^{(1)}$
\begin{widetext}
\bea
\int^{2\pi}_{0} \! \frac{d\vec{q}_\perp}{(2\pi)^2}
\, {\bfG}(\bfq_n,\vec{q}_\perp) &=&
\int^{2\pi}_{0} \! \frac{d\vec{q}_\perp}{(2\pi)^2}
\, \frac{2J_{\perp}}{1+2J_{\perp}\,\(C^{(2)}+{\bf \Sigma}^{(1)}\)} \nn
&=&\int^{2\pi}_{0} \! \frac{d\vec{q}_\perp}{(2\pi)^2}
\left[ \frac{2J_{\perp}}{1+2J_{\perp}\,C^{(2)}} \right.
\left. - \frac{(2J_{\perp})^2 \, {\bf \Sigma}^{(1)}}{\(1+2J_{\perp}\,\(C^{(2)}
+{\bf \Sigma}^{(1)}\)\)
\( 1+2J_{\perp}\,C^{(2)}\)} \right] \nn
&=&{\PV} \! \int^{2\pi}_{0} \! \frac{d\vec{q}_\perp}{(2\pi)^2}
\frac{2J_{\perp}}{1+2J_{\perp}\,C^{(2)}} - 
{\PV} \! \int^{2\pi}_{0} \! \frac{d\vec{q}_\perp}{(2\pi)^2}
\frac{(2J_{\perp})^2 \, {\bf \Sigma}^{(1)}}{\(1+2J_{\perp}\,\(C^{(2)}+{\bf \Sigma}^{(1)}\)\)
\( 1+2J_{\perp}\,C^{(2)}\)} \, .
\eea
\end{widetext}
The first term of the last r.h.s. is the regularized expression for
the RPA-dressed propagator we use in $\Sigma^{(1)}$
whereas the second term (which also ought to be regularized)
is a correction to it at a higher order in $A\,J_\perp/T$.
The prescription scheme consists therefore (at least in this case) in taking
the principal value of the integral (symbolized by $\PV$) which turns out to be finite.


\section{Application to cubic S=1/2 antiferromagnets}
\label{ACL}

Corrections to RPA in the framework of cubic S=1/2 antiferromagnets was
the subject of \cite{Irkhin2000}. In particular, the authors have derived
the integral expression for the diagrams of Fig. (\ref{fig:Hartree_Fock}).
However their correction does not correspond to the first self-energy
correction but rather to a subset of diagrams.
The self-energy correction
derived here includes the re-summation of the one-particle reducible diagrams
made of chains of their contribution.

Let us apply our formalism, using the $\Sigma^{(1)}$ correction to the self-energy
to the compound KCuF$_3$. The experimental value is $T_c=39$ K
\cite{Satija1980}.
The RPA N\'eel temperature is estimated to be $T_c=52.3$ K.
Taking into account their correction, V.Y. Irkhyn and A.A. Katanin then deduced
$T_c=36.7$ K, which correction is of order $30 \%$.
The singularity of the RPA-dressed propagator is removed
by using a semi-empirical approximation due to T. Moriya \cite{Moriya1985}.
But its non-trivial dependence on the temperature and the couplings constant
disappear as well.

From their calculation, one can deduce the value of some intermediate integrals
to be calculated. Making use of the values of these integrals, and therefore resorting
to Moriya's approximation, but within the self-energy correction approach
on obtains the value $T_c=31.2$ K (that is, at this order, re-summing the reducible diagrams).
This correction of order $40 \%$ is significantly stronger.

To determine the critical temperature with the method developed here is
somewhat more complicated. Indeed, the self-energy is also $T$-dependent
in a non-simple way. We have therefore to solve the problem by iteration
on the value of the estimated critical temperature.
The formal calculations detailed above can be applied with
the transverse lattice structure factor
\be
J_\perp(\vec{k}_\perp)=J_\perp(\cos(k_y)+\cos(k_z))
\ee
of three-dimensional cubic lattices.
From Eq. (\ref{landau-ginzburg}) and
after analytic continuation, we obtain
the three-dimensional dynamical magnetic susceptibility
\be
\chi^{\rm xx}_{\rm 3d}(\bfk,\vec{k}_\perp)=\frac{\chi^{\rm xx}_{\rm 1d}(\bfk)
+\Sigma^{\rm xx}(\bfk,\vec{k}_\perp)}
{1-2\,J_\perp(\vec{k}_\perp)\(\chi^{\rm xx}_{\rm 1d}(\bfk)
+\Sigma^{\rm xx }(\bfk,\vec{k}_\perp)  \)} \, .
\ee
The instability condition which can only be satisfied at zero frequency
is therefore
\be
2\, J_\perp (\vec{k})\, X(0,\vec{k})=1 \, ,
\ee
where $X(\bfk)=\chi^{\rm xx}_{\rm 1d}(\bfk)+ \Sigma^{\rm xx}_{(1)}(\bfk)$.
Because at this order $\Sigma^{\rm xx}(\bfk,\vec{k}_\perp)$ does
not depend on $\vec{k}_\perp$ and because it does not a priori
change the monotony of $\chi^{\rm xx}_{\rm 1d}(\bfk)$ with respect to $k_x$,
one can first maximize the l.h.s. on $\vec{k}$.
It leads to N\'eel order in the three directions $k_x=k_y=k_z=\pi$
so that the instability condition is reduced to $4 \, J_\perp \, X(0,\vec{0})+1=0$,
where $J_\perp=J_y=J_z$.

For KCuF$_3$, the exchange values are $J_\parallel=406$ K and $J_\perp=19.1$ K
(5 $\%$ of the main coupling) \cite{Satija1980}.
The small parameter close to the transition is $A\,J_\perp/T \simeq 0.3$.
The numerical result of this calculation is $T_c=40.3$ K, fairly close
to the experimental value.
Finally let us mention that not taking into account the log-log correction would have 
led us to $T_c=37.7$ K.
So the subtle log-log correction would presumably be more significant
than the second order correction $\Sigma^{(2)}(\bfk,\vec{k}_\perp)$.

\begin{figure}[h]
\includegraphics[width=7.5cm]{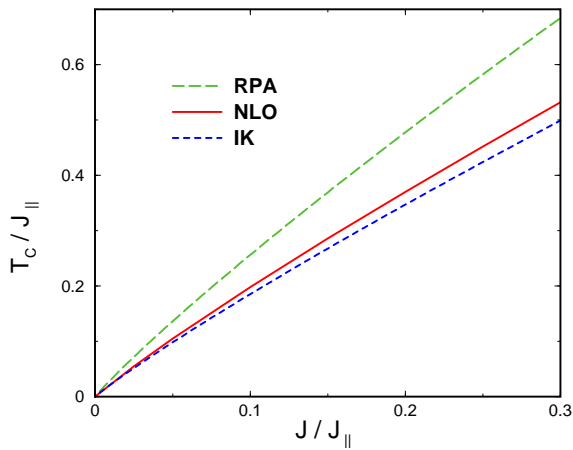} 
\caption{\label{fig:Tc}RPA, next leading order and Irkhin and Katanin'
estimations of the critical temperature $T_c$ of cubic lattices
in units of $J_{\parallel}$ as a function of the interchain exchange
$J=J_{\perp}=J_y=J_z$ in units of $J_{\parallel}$.}
\end{figure}

On Fig.(\ref{fig:Tc}), we have drawn the general curve of the estimated
critical temperature $T_c$ of cubic lattices as a function of the interchain exchange
$J=J_{\perp}=J_y=J_z$. The upper curve (RPA) corresponds to the RPA
estimation of the critical temperature. The lower curve (IK) is deduced from Irkhin and
Katanin' estimation. It can be deduced from their main result \cite{Irkhin2000}
reformulated with our notations and from the use of the exact correlation function prefactor
\be
T_c=k\, J_\perp A\(\frac{\Lambda}{T_c}\) \left[ \frac{\Gamma(1/4)}{\Gamma(3/4)}\right]^2 \,,
\ee
where $k\simeq 0.70$.
The intermediate curve (NLO) corresponds to our next-leading-order estimation
of the critical temperature.
It is significantly lower than the RPA one as expected (about $25 \%$).

\section{Application to frustrated S=1/2 antiferromagnets}
\label{AFL}

\subsection{RPA approach for the dynamical susceptibility in the disordered phase of
Cs$_2$CuCl$_4$}

Cs$_2$CuCl$_4$ is a spin-1/2 frustrated antiferromagnet.
At a temperature of $T_c=0.62$ K, it shows a transition to an ordered phase.
The order is cycloidal. Its parameter is incommensurate and measured to be $k_0=0.186$.
Remarkably, in this phase as well as in the disordered phase,
the excitations spectrum is incoherent \cite{Coldea2000,Coldea1996,Coldea1997a,Coldea1997b}.

Its magnetic Hamiltonian has been recently experimentally determined
with great accuracy \cite{Coldea2001}.
It can be decomposed as~:
\bea
\label{hamiltonien_magnetique}
{\cal H}&=&\sum_k{\cal H}_{\rm plane}^{(k)}+{\cal H}^{(k,k+1)}_{\rm interplane}
+\HDM\ ,\nn
{\cal H}_{\rm plane}^{(k)} &=&J_\parallel\sum_{i,j}
\vec{S}_{i,j,k}\cdot\vec{S}_{i+1,j,k}\nn
&&+J_\perp\sum_{i,j}\vec{S}_{i,j,k}\cdot\left[\vec{S}_{i,j+1,k}
+\vec{S}_{i-1,j+1,k}\right] ,\nn
{\cal H}^{(k,k+1)}_{\rm interplane}
&=& J_z \sum_{i,j}\vec{S}_{i,j,k}.\vec{S}_{i,j,k+1} \, , \nn
\HDM&=& \sum_{i,j,k} \vec{ D}\cdot\left[\vec{S}_{i,j,k}
 \times\vec{S}_{i+1,j,k}\right] \, . 
\eea
In particular a Dzyaloshinskii-Moriya interaction ($\HDM$) has been proven
to exist on the interchain exchange paths, revealed by its anisotropic
nature. 
Experimental estimates for the exchange couplings in ${\rm Cs_2CuCl_4}$
are $J_{\parallel}=4.34$ K, $J_\perp = 1.48$ K (about one third of the main coupling),
$J_z=0.20$ K and finally $|\vec{D}|=0.23$ K (about $5 \%$ of the main coupling)
\cite{Coldea2001,Coldea2000}.

\begin{figure}[h]
\includegraphics[width=6.5cm]{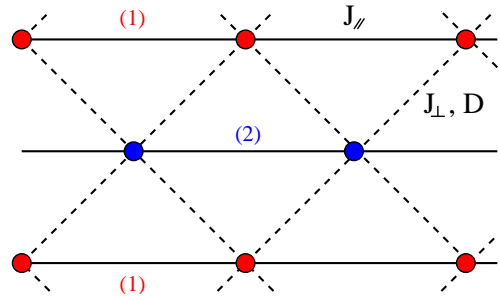} 
\caption{\label{fig:lattice}Exchange paths within the planes~: solid lines denote the strong
exchange $J_{\parallel}$, dashed lines the weaker, frustrated exchange $J_{\perp}$.
Dzyaloshinskii-Moriya exchange of magnitude $D$ also stands on the interchain paths.
Two types of sites among the four in an elementary cell of the lattice
are distinguished.}  
\end{figure}

So it appears that this compound is essentially two-dimensional.
One of the two-dimensional spin lattices is represented on Fig. (\ref{fig:lattice}).
Although the interchain coupling $J_\perp/J_\parallel$ is considerable, the
smallness of the ratio of transition temperature to bandwidth $T_c/\pi
J_{\parallel} \approx 0.05$ indicates that the quasi-one-dimensional
approach we are advocating might be tried on ${\rm Cs_2CuCl_4}$.

An RPA approach has been proven to reproduce qualitative features of the
compound (incommensurability, asymmetry of the dynamical structure factor around
$k_x=\pi$, as well as by construction, the incoherent spectra) \cite{Bocquet2001}.
It also gives very reasonable estimates for the critical temperature, and
the incommensurability.

Given the latest experimental data, and also taking into account the subleading
correction in the logarithmic dependence of the spin correlation functions (which
was intentionally neglected in \cite{Bocquet2001}), RPA gives
$k_0=0.197$ and $T_c=0.85$ K.

Those numbers are off by less than $25 \%$ compared to the experimental values,
which is fairly good given the complexity of the physics displayed by
Cs$_2$CuCl$_4$. It was shown that some qualitative features 
of Cs$_2$CuCl$_4$ were without doubt reproduced by the approach 
advocated in \cite{Bocquet2001}. Knowing the corrections to the RPA result
would hopefully decide whether such a perturbative approach
can be used on this compound for determining quantitative results.

We now briefly reproduce the calculation leading to an RPA-formula for the
dynamical susceptibility as the calculation of the subleading order makes
use of it.

The elementary cell of the Cs$_2$CuCl$_4$ crystal has four Copper ions.
Besides, the vector of the Dzyaloshinskii-Moriya (DM)
interaction which lies on the interchain bounds, points in the third direction
(denoted Oz here).
The orientation of the DM vector is staggered from one plane to another~:
$\vec{D}=\pm D \, \vec{e}_{\rm z}$.

Hence, the magnetic Hamiltonian is now anisotropic.
Those complications make the RPA-dressed propagator possess a matrix-form
which can ultimately (in the complex representation $(+-)$ of the quantum spins)
be reduced to a 4 by 4 matrix.
Eq. (\ref{propagateur_RPA}) is still valid under its matrix form~:
\be
\label{G+-}
\left[{\mathbb G}^{+-} \right]^{-1}={\mathbb C}^{(2)}_{+-}
+ \left[ {\mathbb J}^{+-}_{\perp} \right]^{-1} \, .
\ee
The four components of the vectorial space in which it is defined correspond
to the $-$ component of the four distinct spins in an elementary cell on the right
(resp. $+$ component of the spins on the left).

We have ${\mathbb C}^{(2)}_{+-}=C^{(2)}_{+-} \, {\mathbb I}_4$ and
\be
{\mathbb J}^{+-}_\perp=
\left[
\begin{array}{cccc}
0                       &  {\mathsf J}+{\mathsf K} &  {\mathsf I}            &  0                       \\
{\mathsf J}+{\mathsf K} &                   0      &  0                      &  {\mathsf I}             \\
{\mathsf I}             &                   0      &  0                      &  {\mathsf J}-{\mathsf K} \\
 0                      &          {\mathsf I}     & {\mathsf J}-{\mathsf K} &   0 
\end{array}
\right] \, ,
\ee
where 
\bea
{\mathsf J}(\vec{k}) &=& J_{\perp} \( \cos (k_y) + \cos(k_x-k_y) \) \, , \nn 
{\mathsf K}(\vec{k}) &=& D \(  \sin(k_y) +\sin(k_x-k_y) \) \, , \nn
{\mathsf I}(\vec{k}) &=& J_z\cos(k_z) \, .
\eea
The transverse RPA-dressed propagator of the effective field theory
is related to the transverse RPA imaginary-time correlation function
$\GRPA^{+-}(k_x,\omega)$ (itself related to the transverse RPA dynamical
susceptibility $\chi_{\rm 3d}^{+-}(k_x,\omega)$ by analytic continuation) by
\be
\frac{{\mathbb G}^{+-}}{{\mathbb J}^{+-}_\perp}+{\mathbb J}^{+-}_\perp \GRPA^{+-}={\mathbb I}_4 \, .
\ee
The transverse time-ordered imaginary-time
two-point correlation function of spins is obtained by adding the
contributions from the various sub-lattice correlators, i.e. by taking
e.g.
\be
\sum_{i,j}\langle S^+_{(i)}(\omega,\vec{k})
\, S^-_{(j)}(-\omega,-\vec{k})\rangle \, ,
\ee 
where the summation is over the four types of sites.
After analytic continuation on the frequencies, we obtain the
following RPA expression for the transverse dynamical susceptibility
\bea
\label{chi3d}
&& \chi^{+-}_{\rm 3d}(\bfk,\vec{k}_\perp)= \nn
&& \frac{\chi^{+-}(\bfk)\(1+{\mathsf N}_1(\vec{k})\, 
\chi^{+-}(\bfk)\)}
{\(1-2{\mathsf J}(\vec{k})\, \chi^{+-}(\bfk)+{\mathsf N}_2(\vec{k})
\,\left[\chi^{+-}(\bfk)\right]^2\)} \ ,
\eea
where $\bfk=(\omega,k_x)$ and
\bea
{\mathsf N}_1(\vec{k})&=&{\mathsf I}(\vec{k})-{\mathsf J}(\vec{k})\ ,\nn
{\mathsf N}_2(\vec{k})&=&{\mathsf J}^2(\vec{k})-{\mathsf K}^2(\vec{k})-{\mathsf I}^2(\vec{k}).
\eea
The RPA critical temperature as well as the incommensurability are then determined
through the instability condition obtained by annihilating the denominator of Eq. (\ref{chi3d})
\be
\label{instability}
\({\mathsf J}(\vec{k})\pm \sqrt{{\mathsf K}^2(\vec{k})+{\mathsf I}^2(\vec{k})}\)\chi^{+-}(0,k_x)=1 \, .
\ee
The relevant instability corresponds to the higher possible temperature.
In order to solve Eq. (\ref{instability}) for it,
one can maximize the l.h.s. of  Eq. (\ref{instability}) over $\vec{k}$.
Then one deduces that the instability occurs along the chains direction
$k_y=k_x/2$ and that $k_z=0$ (N\'eel order in the third direction).
Then $k_x$ and $T_c$ have to be determined numerically.

We have assumed that the main instability is given by transverse excitations.
So we need not calculate the longitudinal RPA propagator to calculate the RPA
instability condition.
However, it participates to the next-leading-order correction and we shall need it later.
Eq. (\ref{propagateur_RPA}) is still valid under its matrix form~:
\be
\label{Gzz}
\left[{\mathbb G}^{\rm zz} \right]^{-1}={\mathbb C}^{(2)}_{+-}
+ \left[2\, {\mathbb J}^{\rm zz}_{\perp} \right]^{-1} \, .
\ee
The four components of the vectorial space in which it is defined correspond
to the $z$ component of the four distinct spins in an elementary cell.
We have ${\mathbb C}^{(2)}_{\rm zz}=C^{(2)}_{\rm zz} \, {\mathbb I}_4$ and
\be
{\mathbb J}^{\rm zz}_\perp=
\left[
\begin{array}{cccc}
 0             &  {\mathsf J}  &  {\mathsf I} &  0            \\
{\mathsf J}    &        0      &  0           &  {\mathsf I}  \\
{\mathsf I}    &        0      &  0           &  {\mathsf J}  \\
 0             &  {\mathsf I}  & {\mathsf J}  &   0 
\end{array}
\right] \, .
\ee

\subsection{First correction to RPA}

In the complex spin representation which is more convenient in the case of Cs$_2$CuCl$_4$,
the (matrix) dynamical magnetic susceptibilities are after Eq. (\ref{generatrice_2})~:
\bea
\chi^{+-}_{\rm 3d}(\bfk,\vec{k}_\perp)&=&\frac{\chi^{+-}_{\rm 1d}(\bfk)+\Sigma^{+-}(\bfk,\vec{k}_\perp)}
{1-{\mathbb J}^{+-}_\perp(\vec{k})\(\chi^{+-}_{\rm 1d}(\bfk)+\Sigma^{+-}(\bfk,\vec{k}_\perp)  \)} \, , \nn
\chi^{\rm{zz}}_{\rm 3d}(\bfk,\vec{k}_\perp)&=&\frac{\chi^{\rm{zz}}_{\rm 1d}(\bfk)
+\Sigma^{\rm{zz}}(\bfk,\vec{k}_\perp)}
{1-2\,{\mathbb J}^{\rm zz}_\perp(\vec{k})\(\chi^{\rm{zz}}_{\rm 1d}(\bfk)
+\Sigma^{\rm{zz}}(\bfk,\vec{k}_\perp)  \)} \, .
\eea
Because of the staggering of the DM vector from one plane to another,
there is no chirality on the three-dimensional spin correlation functions
and $\chi^{+-}_{\rm 3d}(\bfk,\vec{k}_\perp)= \chi^{-+}_{\rm 3d}(\bfk,\vec{k}_\perp)$.

\begin{figure}[h]
\includegraphics[width=8cm]{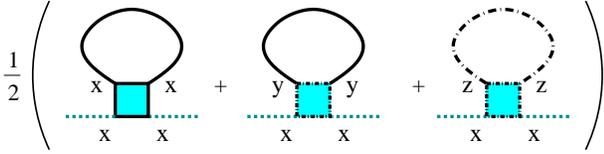} 
\caption{\label{fig:Hartree_Fock_2}Hartee-Fock diagrams with symmetry
factors which are the first
non-trivial terms of the self-energy $\Sigma^{+-}(\omega,\vec{k})$.
The full line propagator is associated with $G^{+-}$, whereas
the dashed-dotted line corresponds to $G^{\rm zz}$.
The full line box is associated with $C^{(4)}_{\rm xxxx}$ whereas the
dashed-dotted line box corresponds to $C^{(4)}_{\rm xxyy}$.}  
\end{figure}

In the particular case of the first correction (Hartree-Fock-like correction), the matrix
$\Sigma^{+-}_{(1)}$ appears to be diagonal. Indeed, it is made up of a single four-point correlation
function which involves four spins belonging to one type of sites (among four).
This does not hold at higher order.
As a consequence, we may see $\Sigma^{+-}_{(1)}$ as a number which is given by~:
\begin{widetext}
\bea
\label{sigma_1_bis}
\Sigma^{+-}_{(1)}\(\bfk ,\vec{k}_\perp\)&=&\frac{A^2}{\beta}
 \sum_n \int_{-\infty}^{\infty} \! \frac{dq_x}{2\pi} \, \left[ \left(
\frac{1}{8} \, C^{(4)}_{\rm xxxx} + \frac{1}{4} \, C^{(4)}_{\rm xxyy} \right)\right.
(\bfk,-\bfk,\bfq_n,-\bfq_n)
\left. \times \int^{2\pi}_{0} \! \frac{d\vec{q}_\perp}{(2\pi)^2}
\, \left[2\, G^{+-}\right](\bfq_n,\vec{q}_\perp)\right] \nn
&&  + \frac{A^2}{\beta}
 \sum_n \int_{-\infty}^{\infty} \! \frac{dq_x}{2\pi} 
\left[ \, \frac{1}{4} \, C^{(4)}_{\rm xxyy} (\bfk,-\bfk,\bfq_n,-\bfq_n) \right.
\left. \times \int^{2\pi}_{0} \! \frac{d\vec{q}_\perp}{(2\pi)^2}
\, G^{\rm zz}(\bfq_n,\vec{q}_\perp) \right] \, .
\eea
\end{widetext}
The three terms in $\Sigma^{+-}_{(1)}(\bfk ,\vec{k}_\perp )$ are derived
from the diagrams on Fig. (\ref{fig:Hartree_Fock_2}).
The integrals over $k_y$ and $k_z$ are performed over an extended Brillouin zone
(from $(k_y,k_y) \in [0,\pi]^2$ to $(k_y,k_y) \in [0,2\pi]^2$) and
the propagators expressions below take this extended scheme into account. 
A similar expression can be obtained for $\Sigma^{\rm zz}_{(1)}$
but is useless for our purpose.

The RPA-dressed propagators $G^{\rm +-}(\bfq_n,\vec{q}_\perp)$
and $G^{\rm zz}(\bfq_n,\vec{q}_\perp)$ can be derived from
Eq. (\ref{G+-})  and Eq. (\ref{Gzz})~:
\bea
G^{+-} &=& \frac{ {\mathsf J}
+ ({\mathsf J}^2-{\mathsf K}^2-{\mathsf I}^2) \, C^{+-}_{(2)} }
{1+2\, {\mathsf J} \, C^{+-}_{(2)} +
({\mathsf J}^2-{\mathsf K}^2-{\mathsf I}^2) \, [C^{+-}_{(2)}]^2 } \, ,\nn
G^{\rm zz} &=& \frac{2\, \( {\mathsf J}+ {\mathsf I}\) }
{1+2\, ({\mathsf J}+{\mathsf I})\, C^{\rm zz}_{(2)} } \, ,
\eea
where $C^{\rm zz}_{(2)}=C^{+-}_{(2)}/2$ is the time-ordered
imaginary-time spin-spin correlation function of the isolated
Heisenberg chains.
When the temperature approaches the theoretical critical
temperature, we expect the second contribution to the self-energy correction
$\Sigma_{(1)}^{+-}$ (which depends on $G^{\rm zz}$) to be 
quantitatively much smaller than the first contribution depending on $G^{+-}$.
Indeed the RPA propagator $G^{\rm zz}$ has an RPA critical
temperature (its pole in $T$) much higher than the one for $G^{+-}$.
It is therefore much less singular that the latter in the temperature
range of interest. This has been checked numerically.

In the case of such a frustrated system, it is less clear what the small
expansion parameter is. However, we can get a rough idea on
inspecting the next-leading-order correction.
First of all there is a prefactor $(A/T)^2$, given by each RPA-line.
In addition there is a dimensionful contribution coming from the transverse
lattice structure factor which depends on the interchain exchange couplings.
But contrary to cubic lattices, it cannot be meaningfully extracted from the
integral.
The RPA-dressed propagator $G^{+-}$ appearing in the integral is more
singular at $(T_c,\vec{k}_0)$ estimated thanks to RPA.
The integral value will therefore be dominated by the value of the
integrands when $\vec{k} \simeq \vec{k}_0$, when $T$ is close to $T_c$.
So, at least when $T$ is close to $T_c$, the expansion parameter is of the
order of $(A\, J_\perp(\vec{k}_0)/T)^2$.
As a consequence, in the case of frustrated quasi-one-dimensional
magnet leading at the transition to an incommensurate order,
$A \, J_\perp/T$ does not necessarily
have to be small provided $A \, J_\perp(\vec{k}_0)/T$ is.
Note that the small parameter in the RPA formula is genuinely
$A \, J_\perp(\vec{k}_0)/T$.

After analytic continuation, we obtain the
following expression for the transverse dynamical susceptibility~:
\be
\label{chi3d_bis}
\chi^{+-}_{\rm 3d}(\bfk,\vec{k}_\perp)= 
\frac{\chi^{+-}_{\rm 1d}(\bfk)\(1+{\mathsf N}_1(\vec{k})\, 
X(\bfk)\)} {\(1-2{\mathsf J}(\vec{k})\, X(\bfk)
+{\mathsf N}_2(\vec{k})
\,\left[X(\bfk)\right]^2\)} \ ,
\ee
with $X(\bfk)=\chi^{+-}_{\rm 1d}(\bfk)+ \Sigma^{+-}_{(1)}(\bfk)$.
The instability condition at this order is therefore
\be
\({\mathsf J}(\vec{k})\pm \sqrt{{\mathsf K}^2(\vec{k})+{\mathsf I}^2(\vec{k})}\)
\, X(0,k_x)=1 \, .
\ee
Because at this order $\Sigma(\bfk,\vec{k}_\perp)$ does not depend on $\vec{k}_\perp$,
it is as easy as in the RPA case to maximize the l.h.s. on $\vec{k}_\perp$.
It leads again to a longitudinal instability and a N\'eel order
from one plane to another.

\subsection{Numerical results}

The numerical computations performed to evaluate the critical
temperature $T_c$ as well as
the incommensurability, are more involved than in the cubic lattice case,
where it is obvious that a N\'eel order prevails below $T_c$.
The instability condition has to be solved with respect to $T_c$ and $k_0$.
The self-energy correction itself, once the obvious dimensionful prefactor
has been put aside, depends on the temperature $T$ through the RPA propagator
and depends on the ratio $k_0/T$ through the four-point correlation function.
An iterative algorithm on $(T_c,k_0)$ can nevertheless be used.
The small parameter of the expansion close to the transition is
$\( -{\mathsf J}+\sqrt{{\mathsf K}^2+{\mathsf I}^2}\)(\pi+k_0) \times A/T \simeq 0.4 \,$.

Our findings are the following.
The critical temperature is estimated to be $T_c=0.66$ K to be
compared to the experimental result $T_c=0.62$ K.
The incommensurability is estimated to be $k_0=0.182$ to be
compared to the experimental result $k_0=0.186$.
This is quite remarkable since the errors for the results obtained
are less than a few percent.

This in return validates the rougher estimates from RPA \cite{Bocquet2001}
which were already quite satisfying. It makes it improbable for the
success of RPA applied to quasi-one-dimensional magnets
to be merely due to chance.


\section{Summary and conclusions}

We have shown that recent one-dimensional exact results from integrable models
and quantum field theory can be applied to one-dimensional spin-1/2
antiferromagnets to compute quantities such as critical N\'eel temperatures.
Their computation can be made systematic in perturbation theory.
On rough grounds, it can be seen as a
high-temperature expansion in $J_\perp/T$.
To the next-leader order, the leading order being the random-phase approximation,
the errors committed differ by less than $10 \%$ from the experimental values
at least in the two cases investigated above.
Although those observables are non-universal, the calculation only depends
on the magnetic Hamiltonian, i.e. exclusively on the knowledge of the
exchange couplings.
Even incommensurate order parameter can be accurately determined this way.
At least this has be shown on the case of the frustrated compound Cs$_2$CuCl$_4$.

The perturbation theory allows more generally to give a perturbative
estimation of the three-dimensional dynamical susceptibility.
But it could as well be used to calculate corrections to multi-spin
three-dimensional correlation functions starting from the one-dimensional functions.

Yet it is somehow hazardous to go beyond the next-leading order mainly used
in this work.
Although the numerical calculations for higher-order correction are achievable,
the resulting correction is likely to be within the field theory approximation error range.
The spin two-point correlation functions are indeed only
asymptotically exact. Being more precise would require to go beyond the knowledge
of the (mathematical) equivalent of the correlations at large distances.
For example, one could include the space-dependent
renormalization group corrections \cite{Barzykin2000,Essler2000}.

{\bf Acknowledgements:} 
The author expresses its gratitude for many helpful discussions with and
constant encouragement  of F.H.L. Essler. Useful discussions with T. Giamarchi,
R. Coldea and D.A. Tennant are also acknowledged.
This work was supported by the EPSRC under grant GR/N19359.


\end{document}